# An Intrusion Detection Architecture for Clustered Wireless Ad Hoc Networks

Jaydip Sen
Innovation Lab, Tata Consultancy Services Ltd.
Bengal Intelligent Park, Salt Lake Electronic Complex
Kolkata, India
e-mail: Jaydip.Sen@tcs.com

*Abstract*—Intrusion detection in wireless ad hoc networks is a challenging task because these networks change their topologies dynamically, lack concentration points where aggregated traffic can be analyzed, utilize infrastructure protocols that are susceptible to manipulation, and rely on noisy, intermittent wireless communications. Security remains a major challenge for these networks due their features of open medium, dynamically changing topologies, reliance on co-operative algorithms, absence of centralized monitoring points, and lack of clear lines of defense. In this paper, we present a cooperative, distributed intrusion detection architecture based on clustering of the nodes that addresses the security vulnerabilities of the network and facilitates accurate detection of attacks. The architecture is organized as a dynamic hierarchy in which the intrusion data is acquired by the nodes and is incrementally aggregated, reduced in volume and analyzed as it flows upwards to the cluster-head. The cluster-heads of adjacent clusters communicate with each other in case of cooperative intrusion detection. For intrusion related message communication, mobile agents are used for their efficiency in lightweight computation and suitability in cooperative intrusion detection. Simulation results show effectiveness and efficiency of the proposed architecture.

## I. INTRODUCTION

A wireless ad hoc network consists of a collection of mobile nodes that communicate with each other through wireless links without the aid of any pre-existing communication infrastructure. Nodes within each other's radio range communicate directly via wireless links, while those that are far apart rely on intermediate nodes to forward their messages. Each node can function as a router as well as a host. Unlike fixed wired networks, wireless ad hoc networks have many operational limitations. For example, the wireless links are constrained by transmission range and bandwidth, and the mobile nodes may have limited battery life, CPU processing power, and memory. The network topology may change rapidly due to mobility of the nodes, and continuous joining and leaving of the nodes in the network. While these characteristics make ad hoc networks more flexible, they introduce security concerns that are either absent or less severe in wired networks. Ad hoc networks are vulnerable to various kinds of attacks that include passive eavesdropping, active interfering, impersonation, and denial-of-service. Although, intrusion prevention measures such as strong authentication and redundant transmission can be used to improve the security of these networks, these techniques can address only a subset of the threats and they are very costly to implement. The dynamic nature of ad hoc networks requires that prevention techniques should be complemented by detection techniques to monitor security status of the network and identify any malicious behavior [1]. Intrusion detection is a second line of defense that provides local security to a node, and also helps in establishing a specific *trust level* of a node in an ad hoc network [2]. Since it is impossible to adopt a fully centralized approach to security in ad hoc networks [3], a cluster-based semi-centralized approach may be adopted that helps in integration of local intrusion detection in a node or in a cluster with network-wide global intrusion detection.

In this paper, we propose an architecture of a cluster-based intrusion detection system for wireless ad hoc networks. In the proposed scheme, an ad hoc network is divided into different clusters using a suitable clustering algorithm [4]. The clustering makes the communication between the nodes in the network more efficient, as each cluster is managed by its cluster-head and inter-cluster communication takes place only through the gateway nodes [5]. The task of cluster management in a cluster is delegated to the cluster-head, which is chosen based on the output an election algorithm that is invoked periodically. The rotation of cluster management responsibility to different nodes ensures a proper load balancing and fault-tolerance in the system [6]. We propose to delegate the cluster-wide intrusion detection responsibility to the cluster-heads, as apart from their default function of cluster management, they can initiate a cooperative approach for intrusion detection. Every node in the network maintains a database of known attacks (misuse signatures). Anomalous activities are defined in terms of upper and lower thresholds for identifying any new attack against the network [7].

The use of mobile agents is proposed for inter-cluster communication. The mobile agents are light-weight and computationally efficient small software components. They enhance the flexibility in cooperative detection ability of a distributed intrusion detection system [8]. However, there have been some security concerns about the mobile agents which need to be investigated further [9][10].

The rest of the paper is organized as follows: Section II presents some related work in the area of intrusion detection in wireless ad hoc networks. Section III describes the architecture of the proposed system. Section IV presents the simulation results carried out on the proposed mechanism. Section V concludes the paper and highlights some future scope of work.

## II. RELATED WORK

Different schemes have been proposed for securing ad hoc networks using intrusion detections schemes [1][8][11]. In a cooperative distributed intrusion detection system proposed by Zhang and Lee [1], every node in an ad hoc network analyzes locally available network data for anomalies. Intrusion attempts are detected by employing a distributed cooperative mechanism. Each node runs intrusion detection agents consisting of six modules. The model uses multi-layer integration approach to analyze the attack scenario. However, the scheme requires large amount of data that needs to be passed over wireless links to update the local database of anomaly and misuse rules. This is certainly a problem in low bandwidth wireless links. Another issue that needs to be addressed is how to obtain enough audit data to establish the normal patterns of users. Without this data, it is almost impossible to carry out anomaly detection accurately.

Li et al. [2] have used mobile agents for developing a coordinated distributed intrusion detection scheme for ad hoc networks. In the proposed scheme, the mobile nodes are divided into different clusters. The cluster-heads act as the manager nodes that contain *assistant mobile agents* and *response mobile agents*. Each cluster-member node (nodes other than the cluster-heads) runs a *host monitor agent* to detect network and host intrusions using *intrusion analyzer* and *interpretation base*. The assistant agent running on a cluster-head is responsible for collecting intrusion-related data from the cluster-member nodes. The response agent on a cluster-head informs the cluster-member nodes about any response initiated by the intrusion detection system against possible intrusive activity on the network. However, the architecture is not modular as there is no separation of functions between the cluster-head nodes and cluster-member nodes. Moreover, it does not use any clustering algorithm to minimize message communication in the network for intrusion detection and response.

Kachriski and Guha [3] have presented an intrusion detection system for ad hoc networks, in which multiple sensors deployed throughout the network collect and merge audit data implementing a cooperative detection algorithm. Sensors are deployed on some of the hosts in the network that monitor the network traffic. The selection of these nodes is based on their connectivity index and the outcome of a distributed voting algorithm. The detection decisions are taken by mobile agents that transport their execution and state information between different sensor hosts of the network, and finally return to the originator host with the result. The authors have proposed two different methods of decision making: independent and collaborative. The approach of independent decision making by mobile agents is susceptible to single point of failure, and therefore, the authors have recommended the use of collaborative approach. The main advantage of this proposition is the restriction of the computation-intensive operations of the system to a few dynamically elected nodes. However, since the mobile agent platforms are themselves vulnerable, the security proposed scheme may be questionable [10].

Albers et al. have proposed a distributed and collaborative architecture of intrusion detection system by using mobile agents [5]. The authors have proposed the use of a local intrusion detection system (LIDS) for monitoring the local activities on each node. Two types of data are exchanged among the LIDS: security data and intrusion alerts. LIDS agents use either the anomaly or misuse detection. Once a local intrusion is detected, the LIDS initiates a response and informs other nodes in the network. Upon receiving an alert, the LIDS protects itself against intrusion by use of a suitable defense mechanism.

Sun et al. have presented an architecture of a *zone-based intrusion detection system* (ZBIDS) that involves a local detection and a collaborative detection technique [12]. The local detection module consists of a general intrusion detection agent model and a Markov chain-based anomaly detection algorithm. To enhance the detection efficiency, the collaborative detection module is utilized. The collaborative detection module works on the ZBIDS agents and uses an aggregation algorithm on the gateway nodes in the clustered ad hoc network. The authors have proposed the IDS for securing routing in the network. The simulation results demonstrate that the proposed scheme is not only efficient in detecting intrusions but also it has reduced false alarm rates appreciably.

Sterne et al. have proposed a dynamic intrusion detection hierarchy that is potentially scalable to large networks [13]. The mechanism is based on a clustering approach, in which the nodes may be organized in a hierarchy with the cluster-head nodes at the top level of the hierarchy. Every node in the network monitors, logs, analyzes, and sends alerts, and responds to the alerts send by other nodes. The cluster heads have the additional tasks of (i) data filtering and data fusion, (ii) detection of intrusions and (iii) security management.

Wang et al. have proposed an end-to-end detection of wormhole attack (EDWA) that is based on a set of mechanisms [14]. In wormhole attack, an adversary builds a tunnel between two end points which are multiple hops way from each other. The message recorded at one is relayed to the other end from where it is broadcasted into the network again. In the proposed defense mechanism against wormhole attack, the authors have proposed a location-based detection mechanism where the source node estimates the minimum hop count to the destination based on the geographic information of the two end hosts in which the receiver's location is piggy-backed by the route reply packet during the route discovery. For, a received route, the source compares the hop count value received from the reply packet with its estimated value. If the received value is less than the estimation, the corresponding route is marked as if a wormhole is detected. The, the source launches wormhole

*tracing* in which the two end points of the wormhole will be identified in a small area provided that there are multi-paths existing between the source and destination. Finally, a normal route is selected for data communication.

III. THE PROPOSED ARCHITECTURE

Our proposed intrusion detection system has a multi-layer cluster-based architecture. At the highest level, the system consists of two broad modules: (i) *Cluster-Head Module* (CHM) and (ii) *Cluster-Member Module* (CMM). These two modules are further subdivided to distribute their specific functionalities. While the CHM runs only on the cluster-head nodes, the CMM runs on all nodes, i.e., the cluster-head nodes and the cluster-member nodes in the network. The intrusion information, detection and response modules are separated with respect to local and global domains. The functionalities of the modules are in the following subsections.

### A. The Cluster Head Module (CHM)

The CHM runs on each cluster-head node, and is responsible for the management of the cluster-member nodes in the cluster. CHM is also responsible for initiating cooperative intrusion detection and response action upon receiving a request from a cluster-member node. The CHM is divided into six modules: (i) cluster management module, (ii) network information module, (iii) mobile agent management module, (iv) global intrusion information module, (v) collaborative intrusion detection module, and (vi) global intrusion response module. The inter-relationships among the modules are depicted in Figure 1. The modules interact with each other via suitable messages in order to collect, store, process, and analyze the data. The functionalities of these modules are described below.

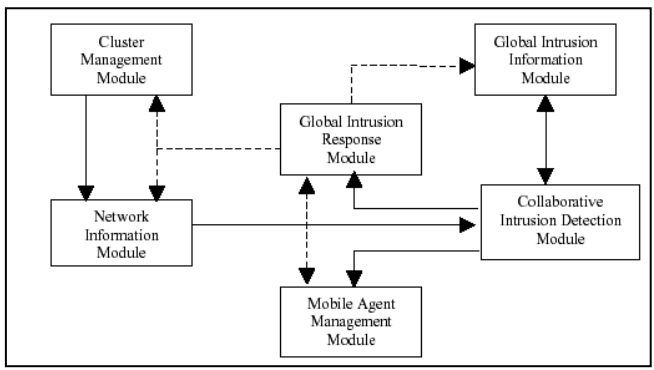

Fig. 1. The architecture of the cluster head module (CHM).
→ represents query messages
---> represents response messages

The *cluster management module* manages the cluster by performing the functions such as: (i) registration of newly joined nodes, (ii) supervision of elections in the cluster, and (iii) communication with other nodes in the cluster for cooperative intrusion detection. This module consists of three sub-modules: (i) cluster-member registration sub-module, (ii) cluster-head election sub-module, (iii) cluster-member communication sub-module. In the rest of this subsection, the functionalities of these modules are described in the rest of this Section.

The *cluster-member registration sub-module* is responsible for managing all the member nodes in its cluster. Every node that comes within the radio range of a cluster-head and becomes a part of the cluster will have to register itself to the cluster-head. The *cluster-head election sub-module* is responsible for managing the elections and successfully forwarding all cluster-related information to the newly elected cluster-head. The *cluster-member communication sub-module* is used for the intra-cluster communication between the cluster-head and the cluster-members. This may be used for cluster-head elections, global intrusion detection, coordination among the mobile agents, local intrusion update, and global intrusion response actions.

The *network information module* keeps track of network-wide information such as information regarding the cluster-head nodes of the neighboring clusters. It is the responsibility of a cluster-head to inform the neighboring cluster-heads about any network-wide intrusion response action, which is further propagated to the cluster members of the neighboring clusters.

The mobile agents are created only at the time of cooperative intrusion detection. The cluster-head of a cluster creates and dispatches the mobile agents and processes the results returned by them. If it detects any intrusion, it informs the nodes in its own cluster and the adjacent clusters for the required response against the intrusion. A database is maintained for the mobile agents that are created and dispatched. Each of the mobile agents is associated with a timer. If a mobile agent returns before the timer expires, the data from the result portion is extracted and its entry is deleted; if the timer expires before the mobile agent returns, the mobile agent is supposed to be lost and a new mobile agent is dispatched. The *mobile agent management module* consists of three sub-modules: (i) mobile agent creation sub-module, (ii) mobile agent dispatch sub-module, (iii) mobile agent deletion sub-module. *Mobile agent creation sub-module* is responsible for creation of the mobile agents when requested by a node for intrusion detection. Mobile agent dispatch sub-module is responsible for establishment of a route for dispatching every mobile agent. *Mobile agent deletion sub-module* destroys a mobile agent when it completes the designated task successfully and returns with the results, or when the timer associated with it expires.

The *global intrusion information module* maintains information about any intrusive activity in a database that is known as the *intrusion interpreter base* [11]. This database is automatically updated whenever any network-wide intrusion is detected. The global intrusion information module consists of two sub-modules: (i) Misuse signature sub-module, and (ii) Anomaly behavior sub-module. While the former contains information regarding all known attack signatures, the latter maintains the definitions of anomalous activities with properly defined upper and lower threshold values. The detection process may use either misuse signature or anomalous activities or both. For power-constrained nodes, only misuse detection may be done.

The *collaborative intrusion detection module* contains the procedures and algorithms that are invoked for cooperative detection of intrusions. This module consists of three sub-modules: (i) mobile agent result processing sub-module, (ii) misuse detection sub-module, and (iii) anomaly detection sub-module. When a mobile agent returns after being processed at all the nodes mentioned in its itinerary, the results that are placed by each node as a result of processing are examined by the cluster-head. The mobile agent result processing sub-module is responsible for processing the results obtained from each node that are collected by the mobile agents. The *misuse detection sub-module* is responsible for detection of intrusions that match with attack signatures existing in the current database. The *anomaly detection sub-module* is responsible for detection of anomalous activities on the basis of suitably defined upper and lower thresholds.

The *global intrusion response module* initiates an action when a cluster member either reports an intrusion, or it detects an intrusion after processing the information collected by the mobile agents. This response action may be local to the cluster or global to the entire network. This module consists of three sub-modules: (i) trust computation sub-module, (ii) cluster-based response sub-module, and (iii) network-wide response sub-module. The *trust computation sub-module* computes the trust for each node. The trust value of a node decreases if it performs any anomalous activity. On the other hand, the trust level gradually increases with time if a node does not perform any anomalous activity. The trust level of a node may only increase till it reaches a maximum value. The *cluster-based response module* initiates an intrusion response after the results from the mobile agents are processed. A cluster-head can also generate a network-wide response. In case of a network-wide response, the *network-wide response sub-module* on a cluster-head node generates a response to the neighboring cluster-heads which, in turn, forward the response in their respective clusters. In this way, the information about a misbehaving node is propagated throughout the network.

*B. Cluster Member Module (CMM)*

The CMM runs on all nodes- cluster-head nodes and the cluster-member nodes in the network. It maintains the data collected locally by the cluster-members about intrusion detection and response. If need arises, it may request the cluster-head node for initiating a cooperative intrusion detection and response action. This happens when the local data available to the CMM are not sufficient to make a concrete conclusion about an intrusive activity and more information from cluster-member nodes in the same cluster or other clusters are required. The CMM is further divided into six sub-modules: (i) cluster-head communication module, (ii) local information module, (iii) mobile agent communication module, (iv) local intrusion information module, (v) multi-layer intrusion detection module, and (vi) local intrusion response module. The inter-relationships among these modules are depicted in Figure 2. The detailed functionalities of these modules are discussed in the following.

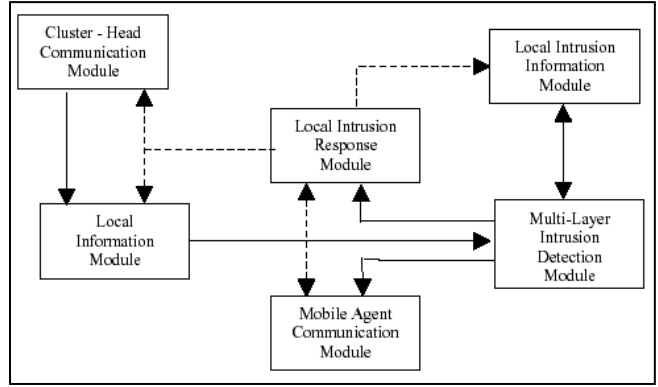

Fig. 2. The architecture of the cluster member module (CMM).
→ represents query messages
--> represents response messages

The *cluster-head communication module* in each node communicates with the cluster-head for sending information regarding a possible local intrusion and also for sending a request to the cluster-head to dispatch mobile agents for initiating a cooperative intrusion detection and response action. This module is divided into two sub-modules: (i) local intrusion update sub-module, (ii) co-operative intrusion detection request sub-module. The *local intrusion update sub-module* informs the cluster-head when it detects any local intrusion in that node. The cluster-head logs the entry and may use this information in future for intrusion detection. The *co-operative detection request sub-module* allows a node to send a request to the cluster-head to initiate cooperative intrusion detection when it cannot conclusively detect an intrusion locally but observes some suspicious activities. On receiving such a request, the cluster-head dispatches a mobile agent to gather information from other members in the cluster, and then processes the gathered information to detect any intrusion in a global scale.

The *local information module* invokes the election algorithm and voting for dynamically electing the cluster-head from the cluster-member nodes. The election algorithm is invoked periodically. When a cluster-head initiates an election, every cluster-member casts its vote for the new cluster-head based on the trust levels associated with the nodes. The node that gets the maximum number of votes is elected as the cluster-head. In case of ties among two or more nodes, other parameters e.g., *connectivity index* [8] may be used to break the tie. The frequency of invocation of the election algorithm depends on the mobility of the nodes in the network. For networks with high mobility, the topology changes fast and election needs to take place frequently. This will ensure that a node elected as the cluster-head will always have a high trust value and a good connectivity index leading to a reduction in message communication overhead among the cluster-member nodes.

The *mobile agent communication module* takes care of the execution of the mobile agents that are dispatched by the cluster-heads. It also collects the results returned by the mobile agents. The module consists of two sub-modules: (i) *mobile agent code execution sub-module*, and (ii) *mobile agent result collection sub-module*. Mobile agent code execution sub-module is responsible for execution of the code contained the mobile agent. The code is executed at each of the nodes in the itinerary of the agent. For a mobile agent to execute, suitable platform must be present in each node. After completion of execution on a node, a mobile agent collects the results and moves to the next node on its itinerary. The *mobile agent result collection sub-module* is responsible for collection of the results of execution. The results returned by a mobile agent are collected and analyzed at the cluster-head from which it originated.

The *local intrusion information module* maintains a database known as 'intrusion interpreter base', which includes the process of *learning* [11]. This database is updated whenever a new intrusion is detected in the network. This module consists of five sub-modules corresponding to the five layers of the TCP/IP stack- application layer, transport layer, network layer, data-link layer, and physical layer. The application layer intrusion sub-module defines the legitimate uses at the application layer such as, the authorized users, their passwords, number of permitted failed login attempts, the applications that can be accessed etc. The transport layer sub-module looks for legitimacy of the established sessions, protocol usage, connection time etc. At the network layer, the important activities are routing table management, route update notification and verification, etc. The channels for communication, the protocol used for communication e.g., CSMA/CD etc are important concerns at the data-link layer sub-module. The physical layer sub-module is concerned with issues related to direct physical access of the system. An attacker can passively eavesdrop on the network traffic as well as the traffic generated by the intrusion detection system, can insert noise in the wireless channel for DoS attack and can also jam the network by injection of spurious messages. The physical layer sub-module must detect these attacks [7].

The *multi-layer intrusion detection module* makes intrusion detection possible at each layer of the TCP/IP stack. It is necessary to detect and respond against intrusions at appropriate layers. As wireless networks cannot have any centralized firewall, intrusion detection in the application layer is necessary for attacks such as 'back-door' or DoS attack [1]. Accordingly, this module consists of five sub-modules corresponding to the five layers of the TCP/IP stack.

The *local intrusion response module* at the each node computes the trust values of the nodes in its cluster and invokes an appropriate intrusion response action, if required. The nodes having high trust values can only participate in election to become a cluster-head node. A cluster member node can take two types of intrusion response actions. On detecting a local intrusion, a node may initiate a response such as blocking the access privilege of the user on the network resources. In this case, it does not communicate this action to the other nodes in the cluster. However, a member node may also communicate its response action to its cluster-head. In that case, the cluster head logs the event and informs other nodes in the same cluster or all the nodes in the network about the event and isolate the offending node from the network.

IV. SIMULATION RESULTS

The proposed scheme has been implemented on network simulator *ns-2* [15] to evaluate its performance. The 802.11 MAC layer in ns2 is used for this purpose. The chosen parameters for simulation are shown in Table I.

Before we discuss the performance results of the system, we describe the simulation for clustering. For cluster formation in the network, we have simulated *passive clustering*.

TABLE I. SIMULATION PARAMATERS

| Parameters | Values |
|---|---|
| Simulation area | 500 * 500 m |
| Number of mobile nodes | 30 |
| Transmission range | 250 m |
| Movement model | Random waypoint |
| Traffic type | CBR (UDP) |
| Channel capacity | 2 Mbps |
| Total number of flows | 15 |
| Avg. packet flow rate | 2 packets/s |
| Packet size | 512 bytes/packet |
| Send buffer at each node | 64 packet (fixed) |
| Training execution time | 1000 s |
| Testing execution time | 50 s |
| Host pause time | 5 s |

Passive clustering is an on-demand protocol. It constructs and maintains the cluster architecture only when there are on-going data packets that piggyback cluster-related information (e.g. the state of a node in a cluster, the IP address of the node etc.). Each node collects neighbor information through promiscuous packet receptions. Passive clustering has also two essential components: (i) first declaration wins rule and (ii) gateway selection heuristic.

With the *first declaration wins rule*, a node that first claims to be a cluster-head, rules the rest of the nodes in its cluster area. Each cluster is assumed to be 2-hop long, i.e., each cluster-member may be at a maximum 2-hop distance from its cluster-head. In passive clustering, to make sure that all the neighbors have been checked, there is no waiting period. This is in contrast to all the weight-driven clustering mechanisms [4]. The cluster-heads are assumed to broadcast their beacons over 2 hops in every 20 seconds time interval.

The *gateway selection heuristic* provides a procedure to elect the minimum number of gateways (including distributed gateways) required to maintain the connectivity in a distributed manner. A gateway is a bridge node that connects two adjacent clusters. The beacon message, sent periodically by a cluster-head in a cluster, contains information that includes the identifications of the cluster-members, and the gateway node in the cluster. The gateway nodes also send beacons to inform the cluster-members about the adjacent clusters. In the proposed scheme, the gateway selection mechanism is designed in such a way that it eventually allows only one gateway for each pair of neighboring cluster-heads. However, in certain situations it may be possible that there is no gateway node between two clusters. This scenario, although very unlikely in reasonably dense ad hoc networks, may occur if two adjacent cluster heads are mutually reachable not by a two-hop route. Then the clustering scheme should select the two intermediate nodes as distributed gateways. Passive clustering maintains clusters using implicit time-out. A node assumes that the nodes it had previously heard from have died or are out of its locality if they have not sent any data within the time-out duration. With a reasonable network communication load, a node can easily keep track of dynamic topology changes by virtue of this time-out.

For the purpose of evaluation of the detection efficiency of the system, we have simulated four types of attacks on the network layer. We have assumed that the goal of the attacker is to degrade the performance of the network or individual nodes instead of gaining privileges of a particular node in the network. This assumption means that the proposed IDS focuses on detecting traffic-related attacks. Some of the well-known attacks in this category are: power exhaustion, storage and CPU exhaustion attacks, network bandwidth exhaustion attacks such as flooding and deprivation attacks, routing-disruption attacks such as blackhole and grayhole attacks etc. [16].

Table II shows the experimental results obtained from the simulation. It is observed that the proposed system have effectively detected the simulated attacks launched against it at the network layer with a very low false positive rates. More sophisticated attack simulations at transport and application layer will be made and results will be reported when available.

TABLE II.    PERFORMANCE RESULTS

| Attack Type | Detection Rate | False Alarm Rate |
|---|---|---|
| Flooding | 100% | 2.8% |
| Blackhole | 99.3% | 0.3% |
| Sleep Deprivation | 90% | 0.7% |
| Packet Dropping (All) | 93% | 0.5% |

V. CONCLUSION

In this paper, we have presented a cluster-based intrusion detection architecture for wireless ad hoc networks. The clustering of the network nodes makes message communication efficient and intrusion detection system robust. Local detection allows for detection of attacks, which are localized to a node or a cluster, whereas global detection involves collaboration among the nodes in different clusters. A mobile agent framework is deployed for communication among the nodes for intrusion related information. The results obtained in simulations show that the scheme is effective and efficient. As a future scope of work, we plan to identify different attack techniques and their consequences at different layers of the TCP/IP stack. We also plan to investigate and determine the optimum number of clusters that maximizes the system performance.